\documentclass[aps,floats,nofootinbib,amssymb,preprint,superscriptaddress]{revtex4}
\usepackage[sort&compress]{natbib}
\usepackage{epsf,epsfig}
\usepackage{amsmath}
\usepackage{hyperref}
\usepackage{subfigure}
\usepackage{graphicx}
\usepackage{color}
\usepackage{mathrsfs}

\newcommand{\be}{\begin{equation}}
\newcommand{\ee}{\end{equation}}
\newcommand{\bea}{\begin{eqnarray}}
\newcommand{\eea}{\end{eqnarray}}
\newcommand{\nn}{\nonumber}
\newcommand{\ba}{\begin{array}}
\newcommand{\ea}{\end{array}}
\newcommand{\bi}{\begin{itemize}}
\newcommand{\ei}{\end{itemize}}

\preprint{
\hbox to \hsize{
\hfill$\vcenter{\hbox{\bf MAD-PH-10-1565}
	\hbox{\bf ANL-HEP-PR-10-61}}$}
}

\everymath{\displaystyle}

\begin{document}

\title{\vspace*{.75in}
Many Leptons at the LHC from the NMSSM}


\author{Vernon Barger}
\affiliation{Department of Physics, University of Wisconsin, Madison, Wisconsin 53706, USA}
\author{Gabe Shaughnessy}
\affiliation{Northwestern University, Department of Physics and Astronomy, Evanston, Illinois 60208, USA}
\affiliation{High Energy Physics Division, Argonne National Laboratory, Argonne Illinois 60439, USA}
\author{Brian Yencho}
\affiliation{Department of Physics, University of Wisconsin, Madison, Wisconsin 53706, USA}

\thispagestyle{empty}

\begin{abstract}
\noindent We present a benchmark in the parameter space of the next-to-minimal supersymmetric standard model (NMSSM) that provides for a dramatic multilepton signal and no jets containing 5 or more leptons resulting from the cascade decays of the third lightest neutralino, $\chi^{0}_{3}$, and the lightest chargino, $\chi^{\pm}_{1}$, via light charged sleptons.  This is a very clean signal with almost no standard model (SM) background.  In some cases, a total signal of $\ge 3~$leptons + 0 jets can be detected at the $5 \sigma$ level at the LHC running at $\sqrt{s}=7$~TeV with approximately 3~fb$^{-1}$ of data and with less than 1 fb$^{-1}$ when running at $\sqrt{s}=14$~TeV.  In addition, kinematic edges in the invariant mass distributions of 2, 3, and 4 leptons are easily detectable with large integrated luminosities ($\sim 600$ fb$^{-1}$) which can lead to simple measurements of the mass differences of heavy particles in the decay chains, including all combinations of the three lightest neutralinos.
\end{abstract}

\maketitle

\section{Introduction}

Multilepton signals are considered one of the best discovery signals of weak-scale supersymmetry at hadron colliders (see, for example, \cite{drees2004,baer2006,titov2007,ball2007,aad2009,cheung2009}  and references therein).  In the case of R-parity conserving supersymmetry, the final state of these decays involve the lightest supersymmetric particle (LSP), typically a neutralino, which then results in large missing energy associated with the multilepton signal.  Many authors have investigated this signal in the context of the minimal supersymmetric standard model (MSSM), often working with a constrained sets of parameters such as minimal supergravity models \cite{baer1995,baer2003,baer2010}, gauge-mediated supersymmetry-breaking models \cite{cheung1997}, or non-universal Higgs masses models \cite{figy2010}. A model independent approach, in which the particle content of the MSSM is used but all mass parameters are taken to be free and independent, was performed in Ref.~\cite{konar2010}.

In these SUSY models, the decay chains are usually initiated by gluino or squark pair production, associated production of squarks and gluinos, or the primary production of a neutralino and chargino pair.  In the latter case, which can be the most relevant when squarks and gluinos are heavy, the largest production cross sections are typically from the process $p p \to W^{\pm} \to \chi^{0}_{2} \chi^{\pm}_{1}$ when gaugino mass unification is assumed \cite{baer2006}.  The second lightest neutralino state, $\chi^{0}_{2}$, is favored here over $\chi^{0}_{1}$ because of its large mixings with the Wino and Higgsino gauge eigenstates, which couple to the $W$-boson.  Both the $\chi^{0}_{2}$ and $\chi^{\pm}_{1}$ then decay down to $\chi^{0}_{1}$ and leptons, neutrinos, and jets via virtual photons, $Z$ and Higgs bosons, sleptons, or squarks.  This is the nature of the multilepton signal derived from the neutralino/chargino sector of the MSSM.

There exist extensions of the MSSM, however, that contain additional neutralino states and therefore allow for the possibility of longer decay chains.  The next-to-minimal supersymmetric standard model (NMSSM) is one well-known example among various singlet-extended models \cite{accomando2006,Barger:2007ay,Barger:2006rd,Barger:2006sk,balazs2007}.  Here, the particle content of the MSSM is extended by one additional gauge-singlet, chiral superfield.  Its effect on the phenemonology of the model is almost entirely due to mixing: the spin-0 singlet $S$ mixes with the Higgs bosons, providing one additional $CP$-even and $CP$-odd state each, while the spin-1/2 singlino $\tilde{S}$ mixes with the neutral fermionic partners of the gauge and Higgs bosons, giving rise to the additional neutralino state.  However, if the mixing is not large it is possible to have a light, singlinolike $\chi^{0}_{1}$ while leaving the composition of the other neutralinos bearing resemblance to the mixings typically seen in the MSSM.  In this case, the dominant sparticle production mode becomes $\chi^{0}_{3} \chi^{\pm}_{1}$, allowing for longer decay chains that may not only result in larger trilepton signals but also provide for multilepton signals with $\ge 5$ leptons.  This scenario was first motivated in Ref.~\cite{barger2006b} and discussed in the context of the constrained NMSSM (cNMSSM) in Ref.~\cite{djouadi2008}.  Similar signals resulting from the NMSSM have also been discussed in Refs.~\cite{ellwanger1998,ellwanger2010,kraml2008}.  We seek to find a benchmark in parameter space that results in such a signal and then to demonstrate its detectability at the LHC.

In Sec.~\ref{sec:model} we discuss the NMSSM, focusing in particular on the neutralino sector of the model and how it is different from the usual MSSM case.  We then describe how we implement this model for the purpose of generating Monte Carlo events in Sec.~\ref{sec:implement} and discuss the chosen benchmark in Sec.~\ref{sec:benchmark}.  We describe how we model LHC detection and the acceptance cuts that we use in Sec.~\ref{sec:cuts}.  The backgrounds to our signal are briefly described in Sec.~\ref{sec:bg} and the detectability of the signal at the LHC and its utility in determining supersymmetric particle mass differences is described in Sec.~\ref{sec:results}.  A final discussion and summary of our results can be found in Sec.~\ref{sec:conclusion}.

\section{NMSSM: An overview}
\label{sec:model}
The NMSSM was introduced as means of alleviating a well-known tension which exists in the MSSM, the so-called ``$\mu$ problem.''  The MSSM superpotential is given by

\be
W_{\rm MSSM} = \mu \hat{H_{u}} \hat{H_{d}} + \hat{u} y_{u} \hat{Q} \hat{H_{u}} - \hat{d} y_{d} \hat{Q} \hat{H_{d}} - \hat{e} y_{e} \hat{L} \hat{H_{d}} ,
\ee

\noindent where $\hat{L}$ and $\hat{Q}$ are the chiral superfields of the lepton and quark doublets, $\hat{e}$, $\hat{u}$, and $\hat{d}$ are the chiral superfields of the lepton, up-type quark, and down-type quark singlets, and $\hat{H_{u}}$ and $\hat{H_{d}}$ are the chiral superfields of the two Higgs doublets \cite{chung2003}.  Here, $\mu$ is the only dimensionful parameter.  It exists in unbroken supersymmetry and is therefore a supersymmetry-conserving parameter that should be of the order of the scale of the complete, unbroken theory.  It is the only such parameter appearing in the Higgs potential; all other dimensionful parameters are soft supersymmetry-breaking parameters that should be ${\cal O}$(TeV).  For the vacuum expectation values (vevs) of the two Higgs states to give $\sqrt{v_{u}^{2} + v_{d}^{2}} = v_{\rm SM} = 246 \; {\rm GeV}$, $\mu$ should itself be of the order of the weak scale.  This creates a naturalness problem.

The solution provided by the NMSSM is to generate the $\mu$ term dynamically by associating it with the vacuum expectation value of a new field.  This is done by removing the $\mu$ term of the MSSM and adding the following two terms to the superpotential:

\be
W_{\rm NMSSM} = W_{\rm MSSM}|_{\mu \to 0} + \lambda \hat{S} \hat{H_{u}} \hat{H_{d}} + \frac{\kappa}{3} \hat{S}^{3} .
\ee

\noindent Here, $\hat{S}$ is a gauge-singlet, chiral superfield and $\lambda$ and $\kappa$ are dimensionless parameters of order unity \footnote{The cubic term forbids a continuous Peccei-Quinn symmetry whose spontaneous breaking would introduce fine-tuning problems associated with bounds on the non-observation of axions .  There remains, however, a discrete $\mathbb{Z}_{3}$ symmetry whose spontaneous breaking introduces a cosmological domain-wall problem.  This is typically circumvented with Planck-suppressed operators that explicitly break the symmetry without affecting the weak-scale phenomenology of the theory (see \cite{maniatis2009} and references therein).}.  Given this superpotential, the effective $\mu$ parameter is then given by

\be
\mu_{\rm eff} = \lambda \left< S \right> = \lambda \frac{s}{\sqrt{2}} .
\ee

\noindent All dimensionful parameters of the neutral scalar potential are now ${\cal O }$(TeV) and the singlet vev, and therefore $\mu$, becomes naturally of the order of the weak scale.

In solving the $\mu$ problem, the addition of a gauge-singlet, chiral superfield to the MSSM alters the mass spectrum of the neutral fields through mixing effects and therefore affects collider phenomenology.  The spin-0 singlet mixes with $H_{u}^{0}$ and $H_{d}^{0}$, the neutral Higgs boson states, to give a total of 3 $CP$-even and 2 $CP$-odd Higgs mass eigenstates.  The effect on the tree level masses, as well as the one-loop corrections derived from the effective Higgs potential, are described in detail in Ref.~\cite{barger2006a}.

For the multilepton signals, we focus our attention on the neutralino and chargino sector of the model.  The neutralino mass matrix is given in the $(\tilde{B},\tilde{W}^{3},\tilde{H}^{0}_{d},\tilde{H}^{0}_{u},\tilde{S})$ basis as

\bea
\textbf{M}_{\chi^{0}} = 
\left( \begin{array}{ccccc}
       M_{1}  &        0     & -g_{1}v_{d}/2          &  g_{1}v_{u}/2          &                      0  \\
       0      &        M_{2} &  g_{2}v_{d}/2          & -g_{2}v_{u}/2          &                      0  \\
-g_{1}v_{d}/2 & g_{2}v_{d}/2 &              0         & -\mu_{\rm{eff}}        & -\mu_{\rm{eff}}v_{u}/s  \\
 g_{1}v_{u}/2 &-g_{2}v_{u}/2 & -\mu_{\rm{eff}}        &              0         & -\mu_{\rm{eff}}v_{d}/s  \\
            0 &            0 & -\mu_{\rm{eff}}v_{u}/s & -\mu_{\rm{eff}}v_{d}/s & \sqrt{2}\kappa s
\end{array} \right) ,
\eea

\noindent where $M_{1}$ and $M_{2}$ are the gaugino mass parameters, $v_{u}$ and $v_{d}$ are the up-type and down-type Higgs vevs such that $v_{u}^{2} + v_{d}^{2} = v_{\rm SM}^{2} = (246 \; {\rm GeV})^{2}$, and $\tan \beta = v_{u}/v_{d}$.   The usual electroweak gauge couplings are $g_{1}$ and $g_{2}$.  The upper-left 4x4 is the standard neutralino mass matrix of the MSSM, while the outer row and column give the singlino contribution.  If the vev $s$ is large and $\sqrt{2} \kappa s < \mu_{\rm eff}, M_{1}, M_{2}$, the lightest neutralino $\chi^{0}_{1}$ can be very singlinolike and light, with a mass $m_{\chi^{0}_{1}} \approx \sqrt{2} \kappa s$.  The effects of such a state, when presumed to be the LSP, have been studied in the context of the dark matter relic density \cite{belanger2005,Barger:2007nv,kraml2008} and collider searches \cite{barger2006b,ellwanger2010}.

The chargino mass matrix is given in the $(\tilde{W}^{\pm},\tilde{H}^{\pm}_{u/d})$ basis as

\bea
\textbf{M}_{\chi^{\pm}} = 
\left( \begin{array}{cc}
M_{2}               & \frac{g_{2}v_{u}}{\sqrt{2}}\\
\frac{g_{2}v_{d}}{\sqrt{2}} & \mu_{\rm eff}
\end{array} \right) .
\eea

\noindent It is left unchanged from the usual MSSM case.  With the exception of the modified mixing matrices, the couplings of the neutralinos and charginos to the $W$-boson and sleptons are also left unmodified.

\section{Model Implementation and Constraints}
\label{sec:implement}
To search for a suitable benchmark and to generate our signal, we extended the standard MSSM implementation included with MadGraph version 4.4.44 \cite{alwall2007}.  To do this, we included the additional neutralino and Higgs boson states and modified all the neutralino and Higgs boson couplings to reflect the additional states and mixings and the effects of direct couplings to the singlet/singlino states.  We independently implement all tree-level sparticle masses and mixing matrices.  In addition, we include the one-loop effective potential corrections to the Higgs boson masses from top / stop loops \cite{barger2006a}.  All decay widths and branching fractions were then calculated using BRIDGE version 2.20 \cite{meade2007} and verified with SPheno version 3.0 \cite{porod2003}, which has hard-coded implementations of the NMSSM branching fractions.

To generate our signal events, we first calculate the two-body process $p p \to W^{\pm} \to \chi^{0}_{3} \chi^{\pm}_{1}$ using our modified MadGraph NMSSM implementation.  Using our own Monte Carlo code, these events are then decayed down to \textit{all possible final states} by using the branching fractions provided by BRIDGE.  These events are then subject to experimental acceptance cuts described below in Sec.~\ref{sec:cuts}.

When searching for a benchmark point, we apply a series of constraints by checking each parameter set using NMSSMtools version 2.3.2 \cite{ellwanger2006}.  This applies basic collider constraints including LEP mass limits, measurements of the anomalous magnetic moment of the muon, and upper limits of $b \to s \gamma$ as well as theoretical constraints, such as the verification of a global minimum of the Higgs potential and the exclusion of points which are found to have Landau poles in $\lambda$, $\kappa$, $h_{t}$, or $h_{b}$ when these couplings are run up to $M_{\rm GUT}$.  NMSSMtools also includes a relic density constraint of $0.094 < \Omega h^{2} < 0.136$, calculated using the NMSSM implementation of Micromegas \cite{belanger2001}.  We are interested in points that do not provide too much dark matter.  Therefore, we only enforce the upper bound $\Omega h^{2} < 0.136$.  In addition to the above constraints applied by NMSSMtools, we perform our own independent checks on the perturbativity of $\lambda$, $\kappa$, and $h_{t}$, the constraints from muon $(g-2)_{\mu}$, and the LEP limits on $ZZH$ couplings.  Our implementations are described in Ref.~\cite{Barger:2007nv}.

\section{Benchmark}
\label{sec:benchmark}

To search for a benchmark point, we note that we are primarily interested in the production mechanism $p p \to W^{\pm} \to \chi^{0}_{3} \chi^{\pm}_{1}$.  There are then in principle several decay chains that can give rise to multilepton signals with $\ge 3$ leptons and no jets.   For example, in the case of heavy sleptons, the decays may be mediated by real or virtual photons and $Z$-bosons:

\be
\chi^{0}_{3} \to V^{(*)} \chi^{0}_{2} \to l^{+} l^{-} V'^{(*)} \chi^{0}_{1} \to  l^{+} l^{-} l'^{+} l'^{-} \chi^{0}_{1} , \\ \nn
\ee

\noindent where $V = A, Z$.  Another interesting possibility is for a hierarchy of the type $M_{\chi^{0}_{3}} > M_{\tilde{l}^{\pm}_{L/R}} > M_{\chi^{0}_{2}} > M_{\tilde{l}^{\pm}_{R/L}} > M_{\chi^{0}_{1}}$.  The charged sleptons will decay as $\tilde{l}^{\pm}_{L/R} \to l^{\pm} \chi^{0}_{i}$ which then allows for the following decay chain:

\be
\chi^{0}_{3} \to l^{\pm} \tilde{l}^{\mp}_{L/R} \to l^{+} l^{-} \chi^{0}_{2} \to l^{+} l^{-} l'^{\pm} \tilde{l'}^{\mp}_{R/L} \to l^{+} l^{-} l'^{+} l'^{-} \chi^{0}_{1} . \\ \nn
\ee

\noindent Similar decays of the chargino are possible if $M_{\chi^{\pm}_{1}}$ is also larger than the charged slepton masses:

\be
\chi^{\pm}_{1} \to \nu_{l} \tilde{l}^{\pm}_{L/R} \to \nu_{l} l^{\pm} \chi^{0}_{2} \to \nu_{l} l^{\pm} l'^{+} \tilde{l'}^{-}_{R/L} \to \nu_{l} l^{\pm} l'^{+} l'^{-} \chi^{0}_{1} .\\ \nn
\ee

\noindent Taken together, this decay chain can lead to signals with up to 7 leptons and no jets.  In order to achieve such a striking signal we seek to generate parameter points that satisfy the following criteria:

\begin{itemize}
\item $\chi^{0}_{3}$ has large Wino and Higgsino components
\item $\chi^{0}_{1}$ is largely singlino ($|Z_{N}^{15}|^{2} > 0.5$) 
\item Charged sleptons are light enough that they mediate the neutralino decays
\end{itemize}

\begin{table}[tb]
\begin{tabular}{ccc} \hline \hline
\multicolumn{3}{c}{Parameter Scan Ranges} \\ \hline \hline
 & Lower Limit & Upper Limit \\ \hline
$M_{1} = \frac{1}{2} M_{2} = \frac{1}{6} M_{3}$ & 25  &  250 GeV \\
$s$ & 2 & 10 TeV \\
$\mu_{\rm eff}$ & $M_{1}$ & 500 GeV \\
$\kappa$ & 0 & $ M_{1}/( \sqrt{2} s )$ \\
$A_{\kappa}$ & -100  & 0 GeV \\
$A_{s}$ & 0 & 1000 GeV \\
$\tan \beta$ & 2 & 10 \\
$A_{t} , A_{b} , A_{\tau}$ & -2000 & 2000 GeV \\
$M_{L} , M_{E}$ & 100 & 200 GeV \\ \hline \hline
\end{tabular}
\caption{Parameter scan ranges used to produce the NMSSM benchmark point.  The soft scalar quark masses are taken to be 2 TeV.  The parameter ranges used here are adapted from Ref.~\cite{belanger2005}.}
\label{tab:scan}
\end{table}

\begin{figure}
\includegraphics[width=0.45\textwidth]{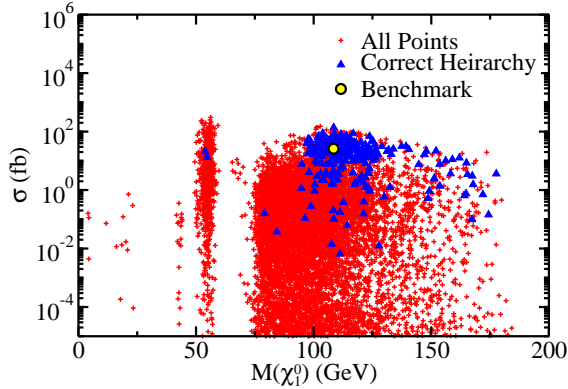}
\caption{The cross section for $p p \to W^{+} \to \chi^{0}_{3} \chi^{+}_{1} \to$~(5 leptons + 0 jets + $X$) versus the mass of the lightest neutralino.  The red plus signs represent all points in the scan, while the blue triangles are for those points which have the mass hierarchy $M_{\chi^{0}_{3}}, M_{\chi^{\pm}_{1}} > M_{\tilde{l}^{\pm}_{L}} > M_{\chi^{0}_{2}} > M_{\tilde{l}^{\pm}_{R}} > M_{\chi^{0}_{1}}$ and also $|Z_{N}^{15}|^{2} > 0.5$.  The yellow circle with the black outline denotes the benchmark point described by Table~\ref{tab:parms}.}
\label{fig:scan}
\end{figure}

\begin{table}[tb]
\begin{tabular}{cccccccccccccccccc} \hline \hline
\multicolumn{17}{c}{Model Parameters} \\ \hline \hline
$\tan \beta$ & $h_{s}$ & $A_{s}$ & $\mu_{\rm eff}$ & $\kappa$ & $A_{\kappa}$ & $A_{t}$ & $A_{b}$ & $A_{\tau}$ & $M_{1}$ & $M_{2}$ & $M_{3}$ & $M_{Q}$ & $M_{U}$ & $M_{D}$ & $M_{L}$ & $M_{E}$ \\ \hline
 7.55 & 0.056 & 488 & 199 & 0.015 &-39.6 & -1170 & 1886 & -143 & 149 & 297 & 891 & 2000 & 2000 & 2000 & 140 & 110 \\ \hline \hline
\end{tabular}
\caption{NMSSM model parameters for the benchmark point.  The dimensionful parameters $\mu_{\rm eff}$, $A$, and $M$ are in GeV.}
\label{tab:parms}
\end{table}

\begin{table}[!htb]
\begin{tabular}{cccccc} \hline \hline
\multicolumn{6}{c}{Sparticle Mass Spectrum (GeV)} \\ \hline \hline
$\chi^{0}_{1}$ :  & 109 & $\tilde{l}^{\pm}_{L}$    : & 147 & $\tilde{u}_{L,R} $ : & 2020 \\
$\chi^{0}_{2}$ :  & 129 & $\tilde{l}^{\pm}_{R}$    : & 118 & $\tilde{d}_{L,R} $ : & 2020 \\ 
$\chi^{0}_{3}$ :  & 191 & $\tilde{\tau}^{\pm}_{1}$ : & 114 & $\tilde{b}_{1} $ : & 2030 \\
$\chi^{0}_{4}$ :  & 206 & $\tilde{\tau}^{\pm}_{2}$ : & 150 & $\tilde{b}_{2} $ : & 2040 \\ 
$\chi^{0}_{5}$ :  & 333 & $\tilde{\nu}_{l}$        : & 125 & $\tilde{t}_{1} $ : & 2010 \\
$\chi^{\pm}_{1}$ : & 173 & $\tilde{\nu}_{\tau}$    : & 125 & $\tilde{t}_{2} $ : & 2100 \\
$\chi^{\pm}_{2}$ : & $\quad $ 333 $\quad$ & & $\qquad \qquad$ & $\tilde{g}$ : & $\quad$ 1060 $\quad$ \\ \hline \hline
\end{tabular}
\qquad \qquad 
\begin{tabular}{cccccc}\hline \hline
\multicolumn{6}{c}{Neutralino Composition} \\ \hline \hline
 & $\tilde{B}$ & $\tilde{W}$ & $\tilde{H}_{u}$ & $\tilde{H}_{d}$ & $\tilde{S}$ \\ \hline
$\chi^{0}_{1}$ : \qquad &     0.02 & $<$ 0.01 &     0.01 &     0.01 &      0.95 \\
$\chi^{0}_{2}$ : \qquad &     0.64 &     0.03 &     0.20 &     0.09 &      0.04 \\
$\chi^{0}_{3}$ : \qquad &     0.33 &     0.17 &     0.26 &     0.24 &  $<$ 0.01 \\
$\chi^{0}_{4}$ : \qquad &     0.01 &     0.01 &     0.47 &     0.51 &  $<$ 0.01 \\
$\chi^{0}_{5}$ : \qquad &     0.01 &     0.79 &     0.06 &     0.14 &  $<$ 0.01 \\ \hline \hline
 & & & & \\ 
\end{tabular} \\ \qquad \\ 
 (a) \hspace{3 in} (b) \\ \qquad \\ 
\begin{tabular}{ccccc}\hline \hline
\multicolumn{3}{c}{Dominant Leptonic Branching Fractions} \\ \hline \hline
$\chi^{0}_{3} \to$ & $l^{\pm }\tilde{l}^{\mp}_{\rm R}$  &  0.40 \\ 
                   & $l^{\pm }\tilde{l}^{\mp}_{\rm L}$  &  0.12 \\
                   & $\nu_{l} \tilde{\nu}_{l} $         &  0.01 \\ \hline

$\chi^{\pm}_{1} \to$ & $l^{\pm} \tilde{\nu}_{l} $         & 0.53 \\
                     & $\nu_{l} \tilde{l}^{\pm}_{\rm L} $ & 0.08  \\ \hline

$\tilde{l}^{\pm}_{L} \to$ & $l^{\pm} \chi^{0}_{2} $ &  0.97 \\ 
                          & $l^{\pm} \chi^{0}_{1} $ &  0.03 \\ \hline

$\chi^{0}_{2} \to$ & $l^{\pm} \tilde{l}^{\mp}_{R}$ & 0.48 \\ 
                   & $\nu_{l} \tilde{\nu}_{l}$ & 0.04\\ \hline 

$\tilde{\nu}_{l} \to$ & $\nu_{l} \chi^{0}_{1}  $ & 1.00 \\  \hline 

$\tilde{l}^{\pm}_{R} \to$ & $l^{\pm} \chi^{0}_{1} $ &  1.00 \\ \hline \hline
\end{tabular}  \\ \qquad \\ (c)
\caption{These tables give, for the chosen benchmark, the (a) mass spectrum of the neutralinos, charginos, sleptons, squarks and gluino in GeV, (b) neutralino composition (mixing elements squared), and (c) leptonic branching fractions.  Here, $l=e,\mu$ and $\tilde{\nu}_{l}$ are the partners of the left-chiral neutrino states.  $\tilde{u}$ and $\tilde{d}$ represent the scalar partners to the first two generations of up-type and down-type quarks, respectively.  The states in (c) are ordered according to descending mass.}
\label{tab:benchmark}
\end{table}

\noindent We proceed with a scan over NMSSM parameter space.  We choose our NMSSM-specific independent parameter set as:

\be
s , \; \kappa , \; A_{\kappa} , \;  A_{s} .
\ee

\noindent We also have the following parameters, which are shared by the MSSM:

\be
\mu_{\rm eff} ,  \;  \tan \beta , \;A_{t} , \;A_{b} , \;A_{\tau}, \; M_{1} , \;M_{2} , \;M_{3} , \;M_{Q_i} , \;M_{U_i} , \;M_{D_i} , \;M_{L_i} , \;M_{E_i} ,
\ee

\noindent where $i=1,2,3$ is a generational index.  We will suppress this index and assume the sfermion mass parameters are the same for each generation.  As we are primarily interested in neutralinos and light slepton superpartners, we set the squark mass parameters to 2 TeV.  We also assume gaugino mass unification: $M_{1} = \frac{1}{2} M_{2} \simeq \frac{1}{6} M_{3}$.

The first condition from the above list suggests that we take $\mu > M_{1}$.  As described in Sec.~\ref{sec:model}, the second condition often arises when $s$ is very large (greater than several TeV) and when $\sqrt{2} \kappa s < \textrm{min}(M_{1},M_{2},\mu)$.  Therefore we take $\kappa < M_{1} / (\sqrt{2} s)$.  Finally, to satisfy the last condition we take $M_{L}$ and $M_{E}$ between 100 and 200 GeV.  The remaining parameter ranges are adapted from the search for a singlinolike LSP satisfying the relic density used in Ref.~\cite{belanger2005}, where $A_{\kappa}<0$ and $A_{s}>0$.   The parameter values and ranges used in our scan are defined in Table \ref{tab:scan}.

We find that large multilepton signals with greater than 5 leptons are fairly generic for parameter points in our scan which have the mass hierarchy

\be
M_{\chi^{0}_{3}}, M_{\chi^{\pm}_{1}} > M_{\tilde{l}^{\pm}_{L}} > M_{\chi^{0}_{2}} > M_{\tilde{l}^{\pm}_{R}} > M_{\chi^{0}_{1}} ,
\label{eq:hierarchy}
\ee

\noindent in addition to a high singlino content of $\chi^{0}_{1}$.  The scan points are exhibited in Fig.~\ref{fig:scan}, which plots $\sigma( p p \to W^{+} \to \chi^{0}_{3} \chi^{+}_{1} \to~ \textrm{5 leptons + 0 jets +}~X )$ versus the mass of the lightest neutralino.  Most points with mass hierarchies described by Eq.~\ref{eq:hierarchy} and with $|Z_{N}^{15}|^{2} > 0.5$ have cross sections that are fairly large ($\sim 1-100$~fb) and which may be detected at the LHC even when including realistic cuts and detector effects.

We choose a single benchmark point from this scan to be used for a more detailed analysis.  The values of all model parameters for this benchmark are given in Table~\ref{tab:parms}.  The resulting mass spectra, neutralino composition, and leptonic branching fractions are given in Tables~\ref{tab:benchmark}~(a)-(c).

\section{Detector Simulation and Acceptance Cuts}
\label{sec:cuts}

We organize our signals according to the number of leptons present and enforce a jet veto on each event.  We do not enforce a $\rlap{\,/}E_{T}$ cut or $\tau$ veto, although these could easily be included.  Since the signal of interest contains leptons and no jets, we perform our analysis on parton-level generated events and use a series of cuts and detector-level effects to roughly simulate actual signal detection.  Our choices for $p_{T}$, $\eta$, and $\Delta R = \sqrt{(\Delta{\phi})^2 + (\Delta{\eta})^{2}}$ cuts are

\be
p_{T} > \left\{
\begin{array}{rl}
20 & \textrm{GeV for the hardest two leptons ($e$,$\mu$)} \\
 7 & \textrm{GeV for all other light leptons} \\
15 & \textrm{GeV for $\tau$ leptons} \\
20 & \textrm{GeV for jets} \\
\end{array}
\right.
\ee

\be
|\eta|  < \left\{
\begin{array}{l}
\textrm{2.4 for electrons} \\
\textrm{2.1 for muons} \\
\textrm{2.5 for $\tau$-leptons and jets} \\
\end{array}
\right.
\ee

\be
\Delta R  >  \left\{
\begin{array}{l}
\textrm{0.2 for light leptons} \\
\textrm{0.4 for all others} \\
\end{array}
\right.
\ee

\noindent Detector smearing of the energy of the jets and leptons is modeled as in Eq.~\ref{eq:smear}.

\be
\frac{\Delta E}{E} = \left\{
\begin{array}{l}
\vspace{0.25cm}
\frac{0.5}{\sqrt{\textrm{$E$/GeV}}} \oplus 0.03 \; \textrm{for jets} \\ 
\frac{0.1}{\sqrt{\textrm{$E$/GeV}}} \oplus 0.007 \; \textrm{for leptons} \\
\end{array}
\right.
\label{eq:smear}
\ee

\noindent We also include basic acceptance cuts and tagging efficiencies according to Ref.~\cite{aad2009} and the effect of isolated leptons from heavy quark decay, which occurs with a probability of $\sim 1/200$ \cite{berger2009}.

We note that both our signal cross sections and the background cross sections described in the next section are calculated at leading order in QCD and do not include the effects of showering or initial and final state radiation.  As we are ultimately interested in signals that do not contain jets at the parton-level and have no colored particles in the primary decay chain, the effects of showering would be minimal and most jets from initial state radiation would presumably be too soft to be tagged with the above criteria.  Initial state radiation would, however, have an effect on the resonant $W$-boson production cross section and kinematics, but as this is a higher-order effect it should not significantly modify our results or conclusions.

\section{Background Analysis}
\label{sec:bg}

\begin{table}[tb]
\begin{tabular}{lcccccccccccc} \hline \hline
\multicolumn{13}{c}{Background Cross Sections (fb)} \\ 
\multicolumn{13}{c}{ $N$~leptons } \\ \hline \hline
& $WZ$ & $ZZ$ & $WWW$ & $WWZ$ & $WZZ$ & $ZZZ$ & $Wt\bar{t}$ & $Zc\bar{c}$ & $Zb\bar{b}$ & $Zt\bar{t}$ & $t\bar{t}$ & TOTAL \\ \hline
$\underline{\sqrt{s}=7~\textrm{TeV}}$ & & & & & & & & & & & & \\ 
$3 l$               &  70 & 7.2  & 0.22 & 0.26  & 0.13 & 0.012 & 1.3   & 5.5 & 5.3  & 1.2   & 7.4  &  99 \\
$\dots$ w/ jet veto &  70 & 7.0  & 0.22 & 0.07  & 0.045& 0.002 & 0.007 & --  &  --  & 0.005 & 1.8  &  80 \\ 
$4 l$               &  -- & 7.2  & --   & 0.07  & 0.005& 0.020 & 0.003 & --  &  --  & 0.12  &  --  &  7.4 \\
$\dots$ w/ jet veto &  -- & 7.2  & --   & 0.06  & 0.003& 0.003 &  --   & --  &  --  & 0.002 &  --  &  7.3 \\
$5 l$               &  -- & --   & --   & --    & --   & --    &  --   & --  &  --  & 0.002 &  --  &  0.002\\
$\dots$ w/ jet veto &  -- & --   & --   & --    & --   & --    &  --   & --  &  --  & --    &  --  &  -- \\ \hline
$\underline{\sqrt{s}=14~\textrm{TeV}}$ & & & & & & & & & & & & \\ 
$3 l$               & 140 & 18  & 0.54 & 1.5   & 0.33 & 0.04 & 3.6  & 19   & 7.5  & 7.7   & 36   & 240   \\
$\dots$ w/ jet veto & 140 & 17  & 0.54 & 0.12  &0.087 & 0.01 & 0.04 & 1.5  &  --  & 0.02  & 3.9  & 170   \\ 
$4 l$               &  -- & 19  & --   & 0.12  &0.027 & 0.01 & 0.01 &  --  &  --  & 0.84  &  --  & 20    \\
$\dots$ w/ jet veto &  -- & 19  & --   & 0.12  &0.027 & 0.01 &  --  &  --  &  --  & 0.013 &  --  & 19    \\ 
$5 l$               &  -- & --  & --   & --    &0.003 & --   &  --  &  --  &  --  & 0.005 &  --  & 0.008 \\
$\dots$ w/ jet veto &  -- & --  & --   & --    &0.003 & --   &  --  &  --  &  --  & 0.003 &  --  & 0.006 \\ \hline \hline
\end{tabular}
\caption{Multilepton background events for the LHC running at $\sqrt{s}=7$ and 14~TeV.  Values are given for each lepton multiplicity before and after the jet veto.  All values include the cuts and detector effects as described in Section \ref{sec:cuts}. }
\label{tab:bg}
\end{table}

The predominant standard model (SM) backgrounds involve the production and decay of weak vector bosons and heavy quarks.  These channels are given in Table~\ref{tab:bg}, along with their corresponding cross sections at the LHC running at $\sqrt{s}=7$ and 14~TeV when cuts and detector effects are taken into account.  The parton-level events were calculated using ALPGEN version 2.13 \cite{mangano2002} while the decays, cuts, and detector effects are later applied.  We note here that for each background involving an on-shell $Z$-boson, there is a corresponding one with a virtual photon that may also contribute.  However, the photon contributions are small relative to those from the $Z$-boson diagrams and we do not include them here.

The largest backgrounds are naturally in the trilepton channel.  These are affected the most by the jet veto, as there are nontrivial contributions coming from isolated leptons originating from heavy quarks in processes such $Zc\bar{c}$, $Zb\bar{b}$, and $t \bar{t}$.  The total backgrounds for four-lepton signals are also nontrivial but could be greatly reduced with a modest cut on $\rlap{\,/}E_{T}$, as nearly all the signal originates from the leptonic decays of $Z$-boson pairs.  Of note is that the backgrounds for final states with greater than 5 leptons are very small, with cross sections of ${\cal O}$(10 ab) or less.

Given these backgrounds with 10~fb$^{-1}$ of data, estimates of the minimum cross sections needed for $3\sigma$ and $5\sigma$ signals in various channels, when including acceptance cuts, are given in Table~\ref{tab:require}.  For simplicity, these are calculated in the Gaussian approximation of signal significance:

\be
\textrm{Significance} = \left(\frac{\sigma_{S}}{\sqrt{\sigma_{S} + \sigma_{B}}} \right) \times \sqrt{\int {\cal{L}} \; \textrm{d} t} \; .
\label{eq:significance}
\ee

\noindent As long as systematic errors on the signal and background cross sections are not large, Eq.~\ref{eq:significance} should provide a reasonable estimate of the reach.

\begin{table}[tb]
\begin{tabular}{cccccccc} \hline \hline
\multicolumn{8}{c}{Required Cross Sections (fb)} \\ \hline \hline
& \multicolumn{3}{c}{$\underline{\sqrt{s}=7~\textrm{TeV}}$} & & \multicolumn{3}{c}{$\underline{\sqrt{s}=14~\textrm{TeV}}$}\\ 
           & $\quad$ $3 l$ $\quad$ & $\quad$ $4 l$ $\quad$ & $\quad$ $5 l$ $\quad$ & & $\quad$ $3 l$ $\quad$ & $\quad$ $4 l$ $\quad$ & $\quad$ $5 l$ $\quad$  \\  \hline
$3 \sigma$ & 8.9  & 3.1  & 0.9  & & 13  & 4.6  & 0.9 \\
$5 \sigma$ & 15   & 5.7  & 2.5  & & 22  & 8.3  & 2.5 \\ \hline \hline
\end{tabular}
\caption{The cross sections for the $N$~lepton + 0 jet signal, after cuts and detector effects, that are necessary for $3 \sigma$ evidence and $5 \sigma$ discovery at the LHC running at 7 and 14 TeV with 10 fb$^{-1}$.}
\label{tab:require}
\end{table}

\section{Results}
\label{sec:results}

\subsection{Signal Rates and Significance}

\begin{table}[tb]
\begin{tabular}{cccccc} \hline \hline
\multicolumn{6}{c}{Signal Cross Sections (fb)} \\ 
\multicolumn{6}{c}{$N$~leptons + 0 jets} \\ \hline \hline
$\sqrt{s}$   & $3 l$ & $4 l$ & $5 l$ & $6 l$ & $7 l$ \\ \hline
      7 TeV  & 25.6  & 4.91  & 2.31 & 0.09 & 0.03 \\ 
     14 TeV  & 68.7  & 13.3 & 6.09  &  0.29 &  0.06 \\  \hline \hline
\end{tabular}
\caption{Multilepton cross sections given by the benchmark NMSSM point for the LHC running at both 7 and 14~TeV, after cuts and detector effects are included.}
\label{tab:signalcs}
\end{table}

\begin{table} [tb]
\begin{tabular}{ccccc} \hline \hline
\multicolumn{5}{c}{NMSSM Signal} \\ 
\multicolumn{5}{c}{$\ge N$ leptons + 0 jets} \\ \hline \hline
 & \multicolumn{2}{c}{$\ge 3~l$} & \multicolumn{2}{c}{$\ge 5~l$}  \\ \hline
 &  $\quad$  Signal $\quad$ &  $\quad$ Background  $\quad$  & $\quad$  Signal $\quad$ &  $\quad$ Background  $\quad$ \\ \hline 
$\underline{\sqrt{s}=7~\textrm{TeV}}$ & & & & \\ 
Cross section (fb) & 33 & 87 & 2.4 & $\sim 0.0$\\ 
Luminosity for $3 \sigma$ (fb$^{-1}$) & \multicolumn{2}{c}{1.0} & \multicolumn{2}{c}{3.7} \\
Luminosity for $5 \sigma$ (fb$^{-1}$) & \multicolumn{2}{c}{2.8} & \multicolumn{2}{c}{10} \\ \hline
$\underline{\sqrt{s}=14~\textrm{TeV}}$ & & & & \\ 
Cross section (fb) & 88.4  & 187 & 6.44 &  0.006\\ 
$N_{\rm events}$ (600 fb$^{-1}$) & $5.3\times10^{4}$ & $1.1\times10^{5}$ & $3.9\times10^{3}$ & 4 \\
Luminosity for $3 \sigma$ (fb$^{-1}$) & \multicolumn{2}{c}{0.32} & \multicolumn{2}{c}{1.4} \\
Luminosity for $5 \sigma$ (fb$^{-1}$) & \multicolumn{2}{c}{0.88} & \multicolumn{2}{c}{3.9} \\ 
\hline \hline
\end{tabular}
\caption{Cross sections for the benchmark signal, and standard model background, in two channels ($\ge 3l$, $\ge 5l$) after accounting for cuts and detector effects for the LHC running at a center-of-mass energies of 7 and 14 TeV.  Also included are the estimated luminosities required for $3\sigma$ and $5\sigma$ discovery.  Here we use $l=e, \mu$.}
\label{tab:discovery}
\end{table}

The cross sections for the signal, at both $\sqrt{s}=$ 7 and 14 TeV, are given for various lepton multiplicities in Table~\ref{tab:signalcs}.  In determining the detectability of the signal at the LHC, we are primarily interested in the signals with $\ge 3$~leptons + 0 jets and $\ge 5$~leptons + 0 jets.  The total rates for both the signal and the background for these processes at $\sqrt{s}=$ 7 and 14 TeV are given in Table~\ref{tab:discovery}.  The approximate luminosities necessary for $3\sigma$ and $5\sigma$ significance are also given in this table and are derived from Eq.~\ref{eq:significance}.

 At the LHC running at 7 TeV, $3\sigma$ evidence can be seen with $\approx 1$~fb$^{-1}$ of data while $5\sigma$ discovery is possible with $\approx 3$~fb$^{-1}$ in the $\ge 3$~leptons + 0 jets channel.  Running at 14 TeV, discovery at the $5\sigma$ level is possible with slightly less than 1~fb$^{-1}$ of data when considering the $\ge 3$~leptons channel.  While the $3l$ channel is not unique to the NMSSM as it may be a discovery channel for the MSSM, the $5l$ channel is more exclusive.  The LHC may observe the $5l$ channel with as little as 10 fb$^{-1}$ of data running at 7 TeV and 4 fb$^{-1}$ of data running at 14 TeV.  While nonobservation of this signal would not allow the NMSSM as a whole to be excluded, the mass hierarchy described in Sec.~\ref{sec:benchmark} would typically give rise to such a signal.

\subsection{Kinematic Edge Measurements}

\begin{figure}[tb]
\subfigure[]{\includegraphics[clip,width=0.45\textwidth]{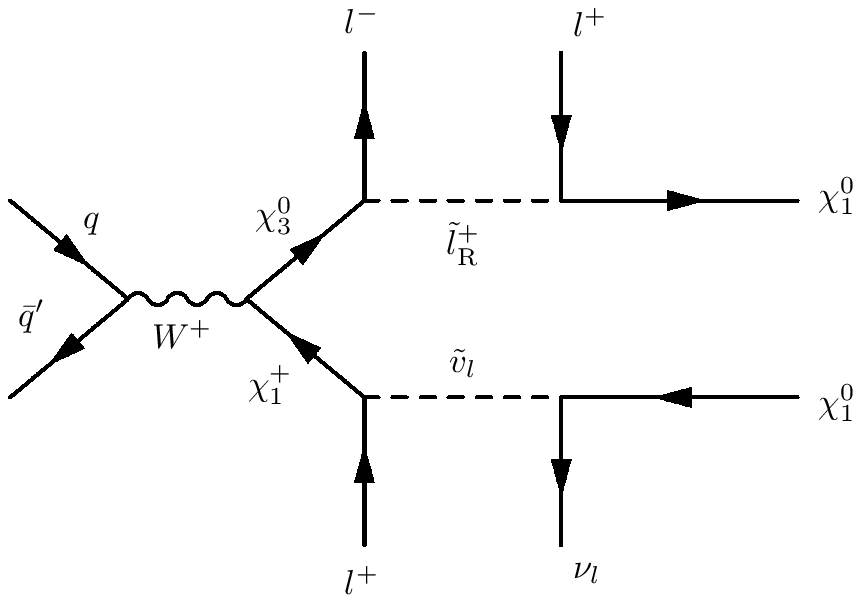}} \qquad
\subfigure[]{\includegraphics[clip,width=0.45\textwidth]{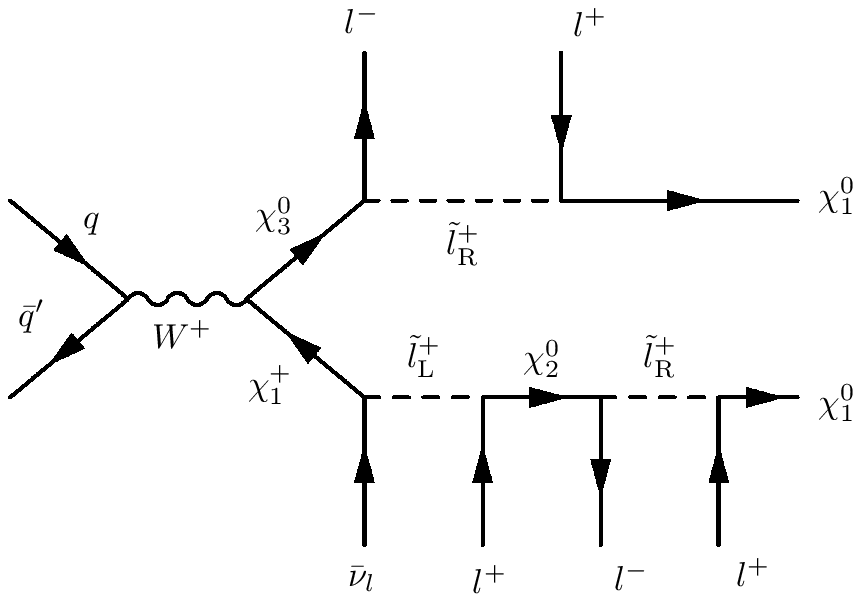}}
\subfigure[]{\includegraphics[clip,width=0.45\textwidth]{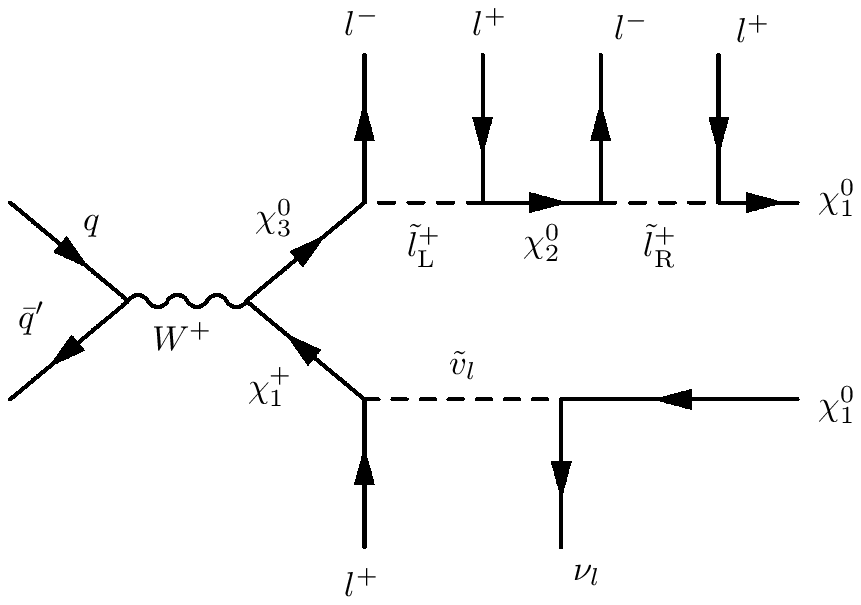}} \qquad
\subfigure[]{\includegraphics[clip,width=0.45\textwidth]{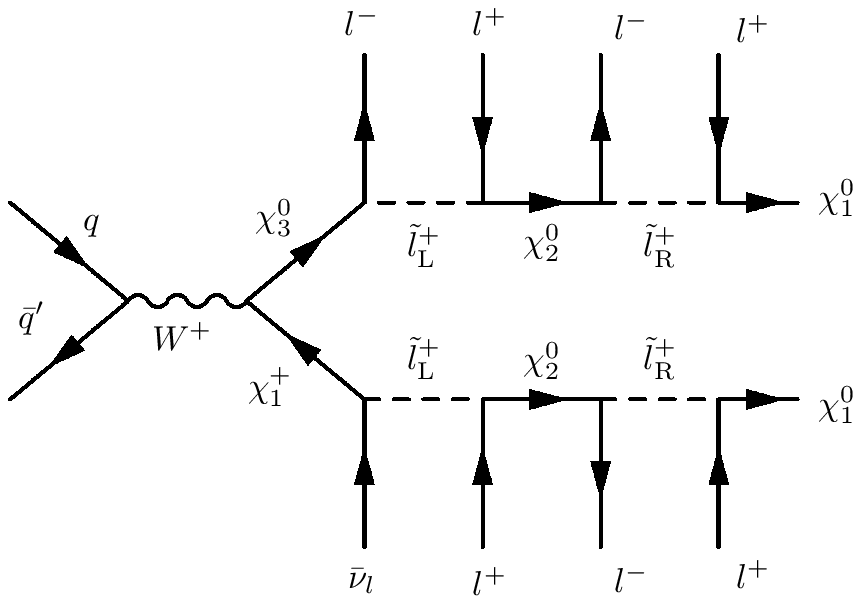}}
\caption{Primary diagrams contributing to (a) 3 lepton final states (b)/(c) 5 lepton final states and (d) 7 lepton final states.}
\label{fig:decays}
\end{figure}

\begin{table}[tb]
\begin{tabular}{cccc} \hline \hline
Process &  $\;$ Total Mass Difference (GeV) & $\;$ Kinematic Edge (GeV) & $\quad$ Invariant Mass Distribution $\quad$ \\ \hline \hline

$\chi^{0}_{3} \to \chi^{0}_{2}$                  & 62.4 & 59.3 & 2L-OS-SF   \\
$\chi^{0}_{3} \to \chi^{0}_{1}$                  & 82.5 & 82.5 & 4L  \\ 

$\chi^{0}_{3} \to \tilde{l}^{\pm}_{R}$           & 72.6 & 72.4 & 3L  \\ \hline 

$\tilde{l}^{\pm}_{L} \to \tilde{l}^{\pm}_{R}$    & 28.9 & 28.0 & 2L  \\ 
$\tilde{l}^{\pm}_{L} \to \chi^{0}_{1}$           & 38.8 & 38.8 & 3L  \\ \hline

$\chi^{0}_{2} \to \chi^{0}_{1}$                  & 20.1 & 20.1 & 2L-OS-SF  \\ \hline \hline

\end{tabular}
\caption{Various subprocesses involved in the decay of $\chi^{0}_{3}$ as in Fig.~\ref{fig:decays}(d).  The total mass difference between the initial and final heavy particles of each process are given along with the true kinematic upper limit in the associated invariant mass distribution.  In this table, $l=e, \mu$.  The masses and mass differences are given in GeV.  The distributions are labeled by the number of leptons and, in the case of lepton pairs, whether they are required to be opposite-sign (OS) and same-flavor (SF).}
\label{tab:edges}
\end{table}

\begin{figure}[tb]
 \qquad \\
 \qquad \\
\subfigure[]{\includegraphics[clip,width=0.45\textwidth]{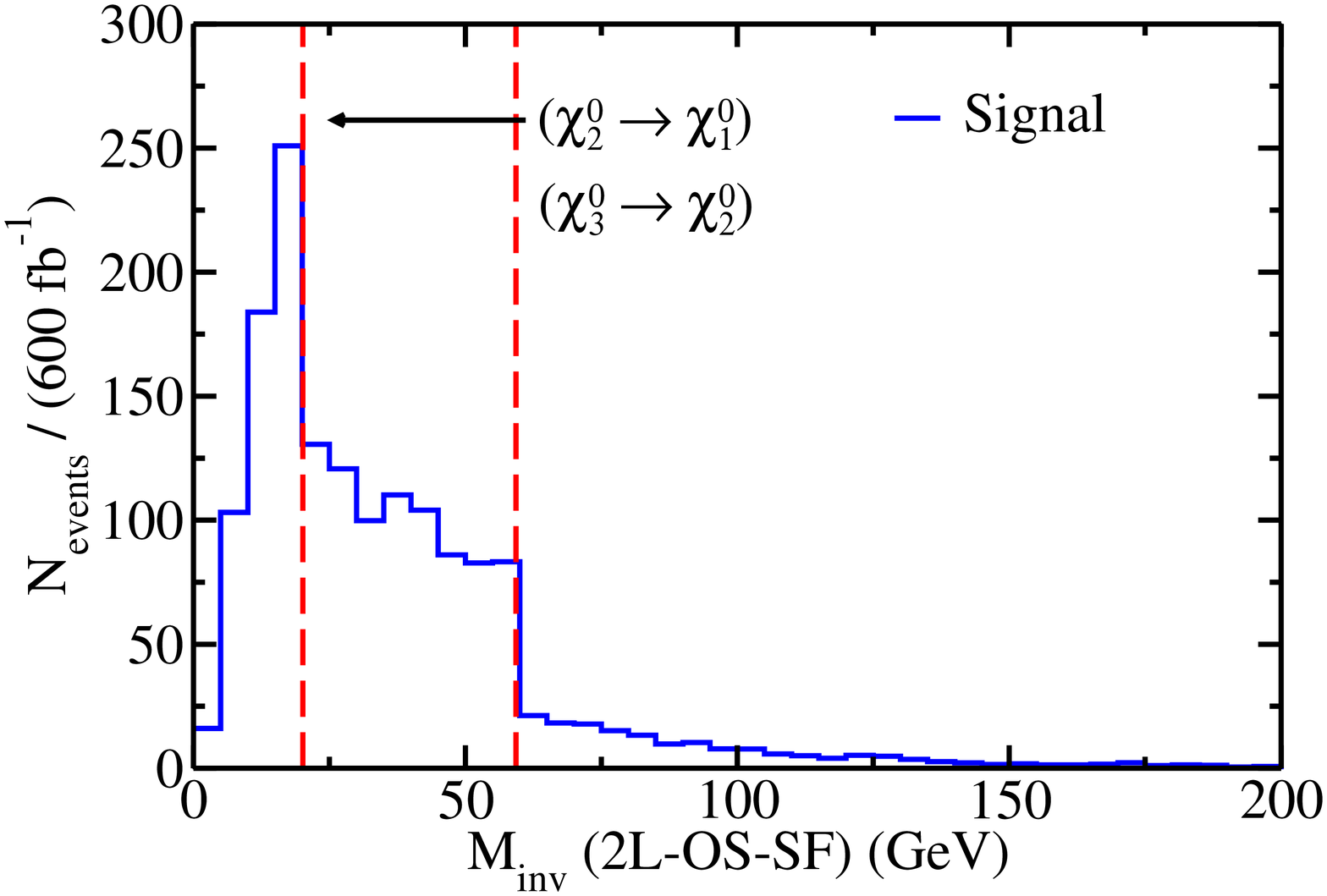}}
\subfigure[]{\includegraphics[clip,width=0.45\textwidth]{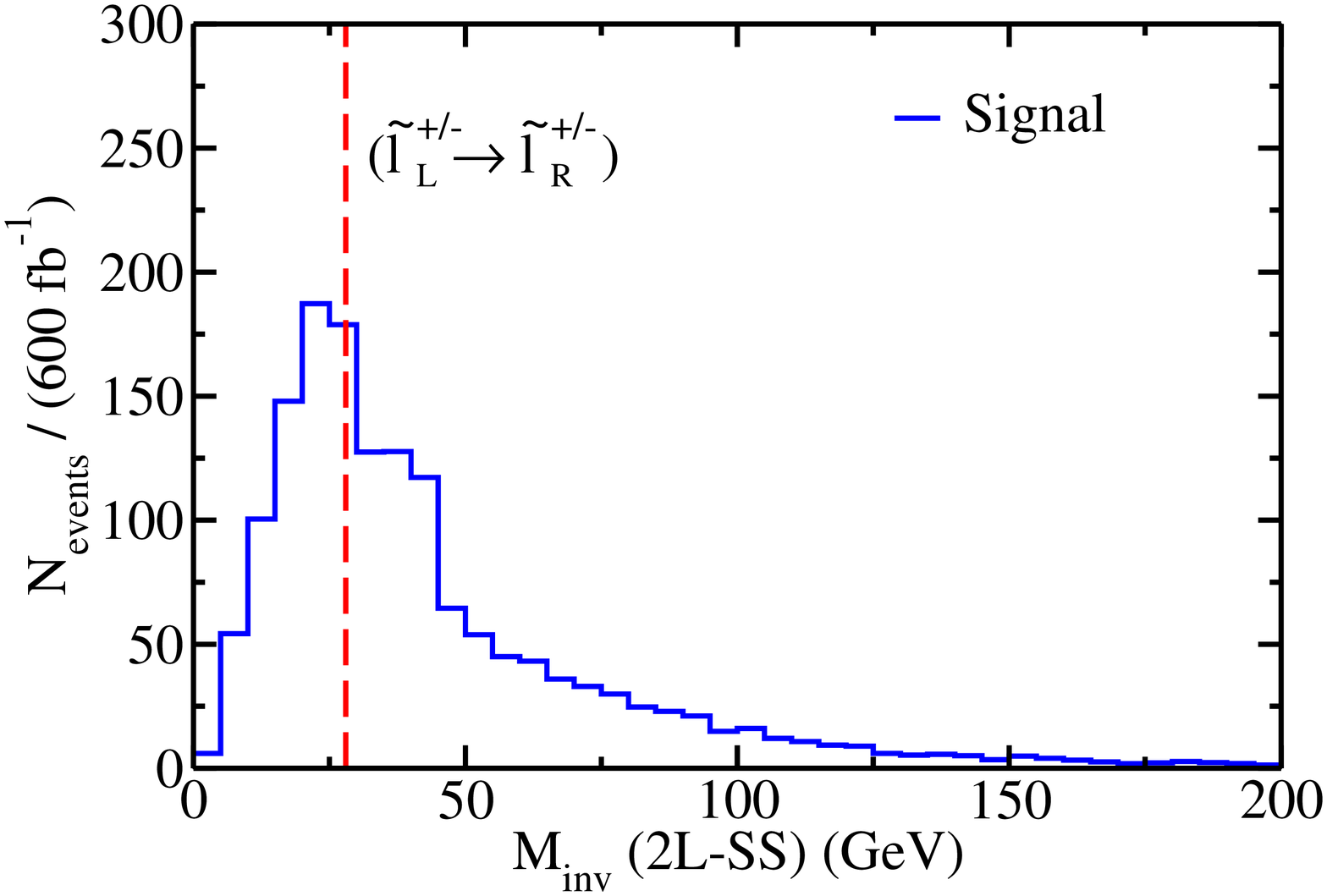}}
\subfigure[]{\includegraphics[clip,width=0.45\textwidth]{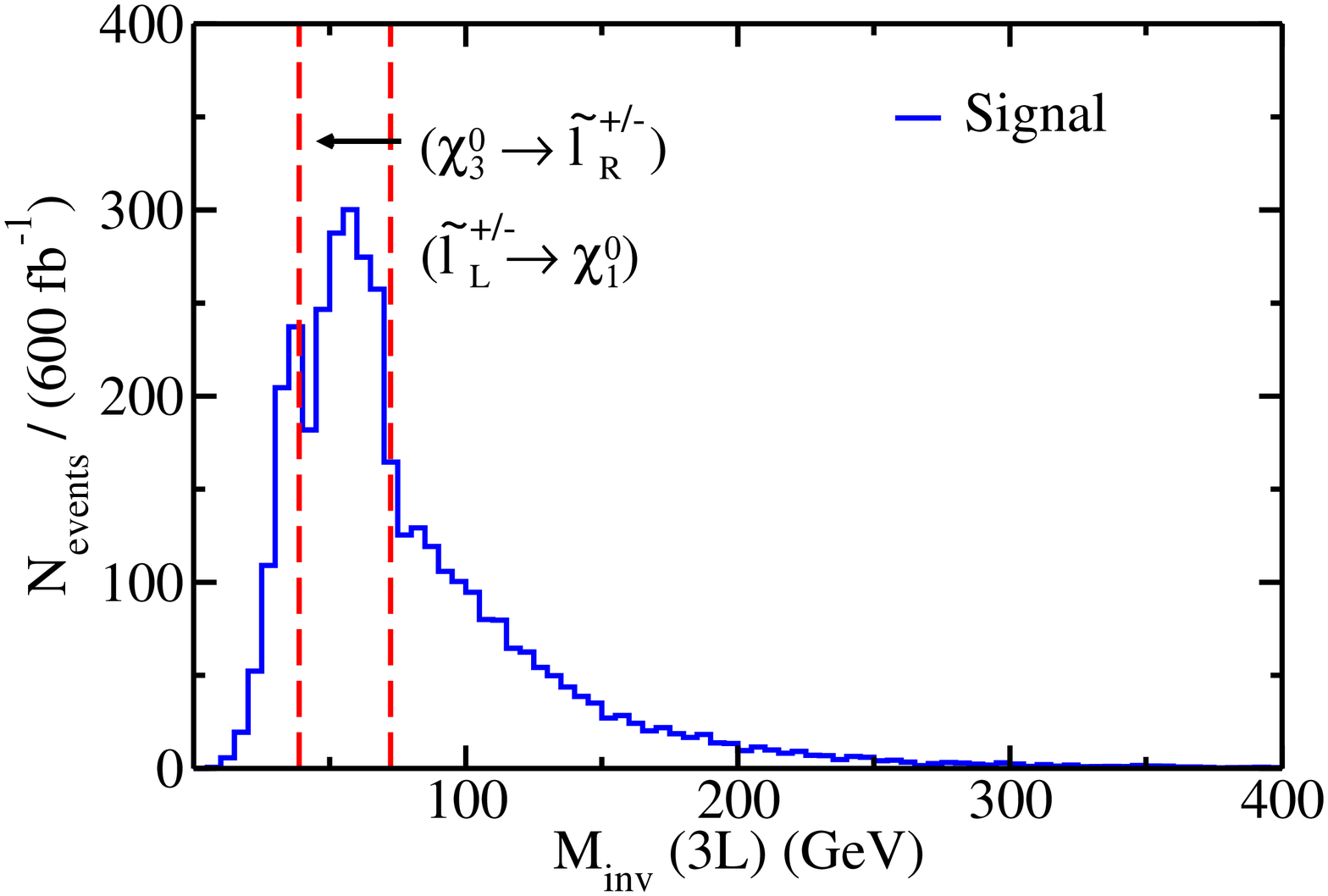}}
\subfigure[]{\includegraphics[clip,width=0.45\textwidth]{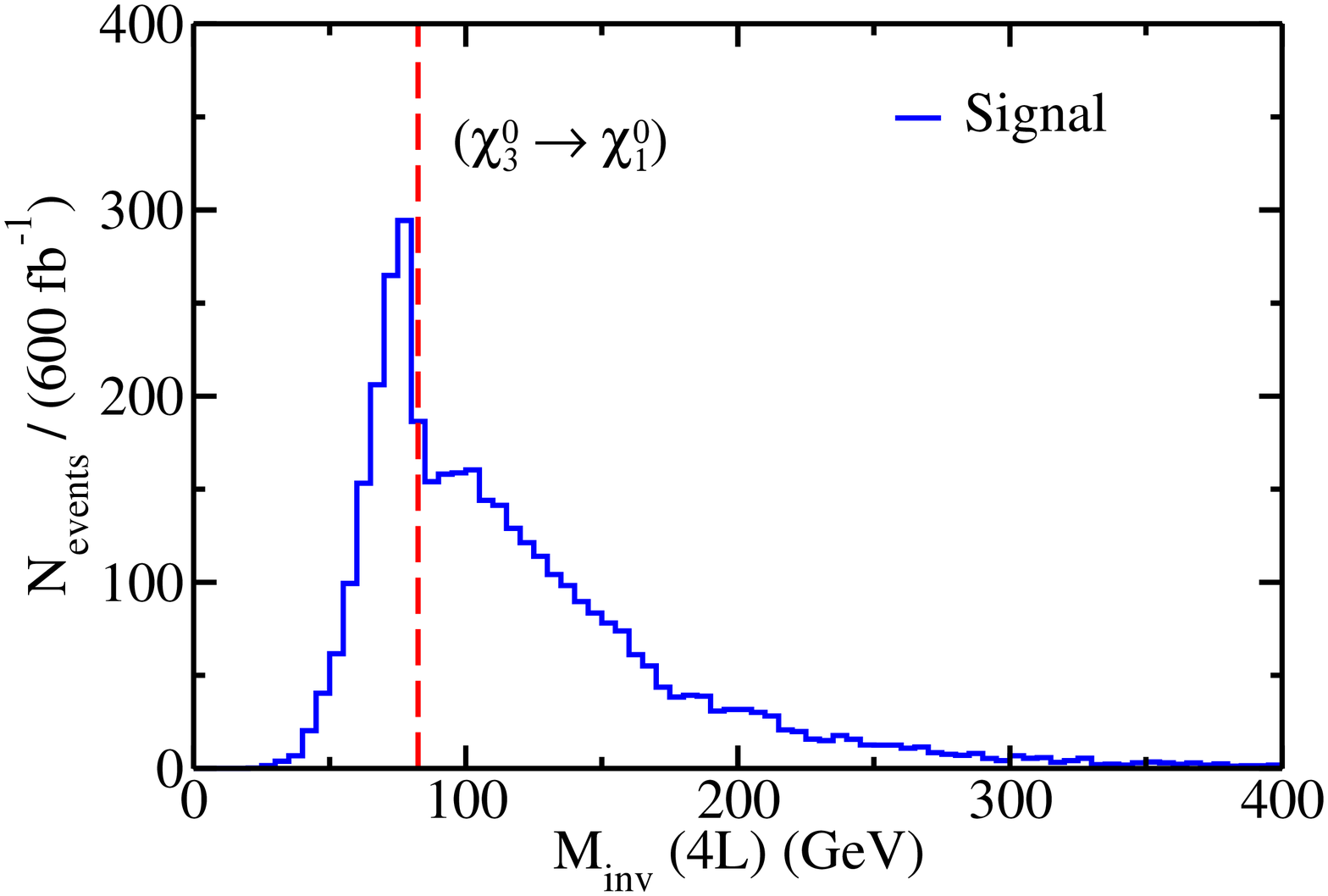}}
\caption{Invariant mass distribution of dileptons of (a) opposite-sign, same-flavor leptons and (b) leptons of the same sign and for (c) three and (d) four leptons.  These distributions are all derived from the $\ge 5$~leptons + 0 jets channel at the LHC running at $\sqrt{s}=$14~TeV with 600 fb$^{-1}$ of integrated luminosity.  In all channels, the backgrounds are negligible.  The vertical dashed lines indicate the kinematic upper limits for various subprocesses which are labeled by the initial and final heavy particles involved.  In these plots, $l=e, \mu$.} \qquad \\
\label{fig:minv}
\end{figure}

Given such a large multilepton signal, it is then possible to estimate several mass differences from the kinematic edges that will be observable in the invariant mass distributions of the leptons when enough integrated luminosity is accumulated \cite{bachacou1999,Gjelsten:2005aw,Lester:2005je,Cheng:2008mg,Burns:2009zi,Cheng:2009fw,Barr:2010zj,Han:2009ss}.  Analytic expressions for the kinematic edges in the invariant mass distributions of 2, 3, and 4 leptons resulting from a sequence of two-body decays are given in Appendix~\ref{app:edges}.  For a decay chain involving a sequence of two-body decays, each emitting one massive particle and one massless lepton, the absolute upper limit of the invariant mass of the leptons is given by $M_{\rm max}(n~\textrm{leptons}) = (m_{I} - m_{F})$, where $m_{I}$ is the mass of the initial mother particle and $m_{F}$ is the mass of the heavy particle resulting from the final two-body decay of the chain.  While this limit can not always be saturated, it is a fairly good approximation to the true upper limit when this mass difference is not too large compared to the masses involved.  We can therefore interpret these kinematic limits as probes of mass differences of particles at various steps in the decays of $\chi^{0}_{3}$ and $\chi^{\pm}_{1}$.

The primary decay chains giving rise to our multilepton signals are illustrated in Figs.~\ref{fig:decays}~(a)-(d).  Each step in the decay of $\chi^{0}_{3}$ results in the emission of light charged leptons via two body decays.  Every combination of 2, 3, and 4 leptons in adjacent steps of the decay chain therefore offers the potential to measure mass differences by means of kinematic edges.  The mass differences that may be probed and the true kinematic limits of the associated invariant mass distributions are summarized in Table~\ref{tab:edges}.  It is clear that, in most cases, $M_{\rm max}(n~\textrm{leptons}) = (m_{I} - m_{F})$ is a good approximation to the true value.  The most useful tool is the invariant mass of opposite-sign (OS) and opposite-flavor (OF) lepton pairs, which can be used to measure the mass differences $M_{\chi^{0}_3}-M_{\chi^{0}_2}$ and $M_{\chi^{0}_2}-M_{\chi^{0}_1}$.  It can also be used to measure the mass difference $M_{\tilde{l}^{\pm}_{L}} - M_{\tilde{l}^{\pm}_{R}}$.  However, this difference is not limited to opposite-sign, opposite-flavor lepton pairs; the Majorana nature of the neutralinos allows their decay into leptons of either charge and flavor and the same is true for the decays of the charged sleptons to neutralinos.  One of the best ways to measure this edge is then to use same-sign dileptons.  Beyond the dilepton invariant mass distributions, the invariant mass distribution of 3 leptons can be used to measure the differences $M_{\tilde{l}^{\pm}_{L}} - M_{\chi^{0}_{1}}$ and $M_{\chi^{0}_{3}}- M_{\tilde{l}^{\pm}_{R}}$, while 4 leptons may be used to measure $M_{\chi^{0}_{3}} - M_{\chi^{0}_{1}}$.

The decay of $\chi^{\pm}_{1}$ offers fewer opportunities to measure differences involving its mass; there will always be a neutrino emitted in either the first or second step of the decay chain, resulting in a loss of crucial kinematic information.  However, when $\chi^{\pm}_{1} \to \nu_{l} \tilde{l}^{\pm}_{L}$, the subsequent decay of the slepton will provide information as described above.

To demonstrate these effects, we look at various invariant mass distributions of leptons in the $\ge 5$~leptons + 0 jet signal with 600 fb$^{-1}$ of data \footnote{We choose such a large integrated luminosity to illustrate the kinematic edge effect in the absence of large statistical fluctuations in the data.} at the LHC running at $\sqrt{s}=14$~TeV.  While offering less signal significance than the case of $\ge 3~$leptons due to a smaller overall rate, the virtual lack of background makes this signal ideal for kinematic edge searches.  We plot the dilepton invariant mass distributions for the $\ge 5$~leptons + 0 jet signal in Figs.~\ref{fig:minv}~(a)-(b) and the three- and four-lepton distributions in Figs.~\ref{fig:minv}~(c)-(d).  As mentioned above, there is essentially no background to these signals and each kinematic edge listed in Table~\ref{tab:edges} can be clearly identified.  Therefore, not only is the multilepton + 0 jet channel excellent for the discovery of this type of NMSSM benchmark, it can easily provide important mass information as well.

\section{Conclusions}
\label{sec:conclusion}

We have investigated a scenario in NMSSM parameter space in which a light, singlinolike neutralino is the LSP.  This allows for extended decay chains from the production $ p p \to W^{\pm} \to \chi^{0}_{3} \chi^{\pm}_{1}$ which can then result in large multilepton signals of high lepton multiplicity.  In particular, we find that signals with $\ge 5$ leptons and 0 jets in the final state can be quite large given the mass hierarchy $M_{\chi^{0}_{3}}, M_{\chi^{\pm}_{1}} > M_{\tilde{l}^{\pm}_{L}} > M_{\chi^{0}_{2}} > M_{\tilde{l}^{\pm}_{R}} > M_{\chi^{0}_{1}}$, a singlinolike LSP, and heavy squarks.  The backgrounds for such processes are virtually negligible at the LHC.  Therefore, when considering a representative benchmark point and looking at a signal with $\ge 3$ leptons and 0 jets in the final state, $5\sigma$ discovery is possible at the LHC running at $\sqrt{s}=7$ TeV for approximately 3~fb$^{-1}$ and it is possible with less than 1~fb$^{-1}$ of data when running at 14 TeV.  In addition, the $\ge 5l$ + 0 jet channel, which is more unique to this benchmark, can be discovered at the $5\sigma$ level with 10 fb$^{-1}$ of data running at 7 TeV and 4 fb$^{-1}$ of data running at 14 TeV.

The high multiplicity multilepton signals are also quite useful for measuring mass differences by looking at kinematic mass edges in large amounts of accumulated data.  When looking at the invariant mass distributions of two, three, and four leptons in the $\ge 5$ leptons + 0 jets signals with 600~fb$^{-1}$ of data, these kinematic edges are clearly visible, allowing for the determination of nearly every mass difference present in the decay of $\chi^{0}_{3}$.  High multiplicity multilepton signals are therefore a very useful tool to use at the LHC should this particular scenario of the NMSSM be realized.

\bigskip
{\bf Acknowledgments}

The authors would like to thank the referee for the constructive comments that were given and W. Zhu and P. Langacker for their valuable early participation in this study.  This work was supported in part by the U.S. Department of Energy under Grant Nos.~DE-FG02-95ER40896, DE-AC02-06CH11357, DE-FG02-91ER40684 and in part by the Wisconsin Alumni Research Foundation.

\newpage

\appendix
\section{Kinematic Edge Expressions}
\label{app:edges}

For the decay chain $A \to b B \to b c C$, where each step represents a two-body decay and $b$ and $c$ are massless particles, the kinematic upper limit for $M^{2}(b c)$ is given by

\be
M^{2}_{\rm max}(b c) = \frac{(m_{A}^{2}-m_{B}^{2}) (m_{B}^{2}-m_{C}^{2})}{m_{B}^{2}}  \le (m_{A} - m_{C})^{2} ,
\ee

\noindent where $m_{A}$, $m_{B}$, and $m_{C}$ are the masses of particles $A$, $B$, and $C$, respectively \cite{bachacou1999}.

For the decay chain $A \to b B \to b c C \to b c d D$, where $b$, $c$, and $d$ are massless particles, the kinematic upper limit for $M^{2}(b c d)$ is given by

\begin{equation}
M^{2}_{\rm max}(b c d) = \left\{
\begin{array}{lcccc}
\vspace{0.25cm}
\frac{(m_{A}^{2}-m_{B}^{2}) (m_{B}^{2}-m_{D}^{2})}{m_{B}^{2}} & \textrm{iff} & & \frac{m_{A}}{m_{D}} > & \frac{m_{B}^{2}}{m_{D}^{2}} \\
\vspace{0.25cm}
\frac{(m_{A}^{2} m_{C}^{2}- m_{B}^{2} m_{D}^{2}) (m_{B}^{2}-m_{C}^{2})}{m_{B}^{2} m_{C}^{2}} & \textrm{iff} &  & \frac{m_{A}}{m_{D}} < & \frac{m_{B}^{2}}{m_{C}^{2}} , \\
\vspace{0.25cm}
\frac{(m_{A}^{2}-m_{C}^{2}) (m_{C}^{2}-m_{D}^{2})}{m_{C}^{2}} & \textrm{iff} &  & \frac{m_{A}}{m_{D}} < & \frac{m_{C}^{2}}{m_{D}^{2}} \\
\vspace{0.25cm}
(m_{A}-m_{D})^{2} & \textrm{otherwise} & & & \\
\end{array} \right.
\end{equation}

\noindent where $m_{A}$, $m_{B}$, $m_{C}$, and $m_{D}$ are the masses of particles $A$, $B$, $C$, and $D$ \cite{Lester:2005je,baer2006,Burns:2009zi}.

For the decay chain $A \to b B \to b c C \to b c d D \to b c d e E$, where  $b$, $c$, $d$, and $e$ are massless particles, the kinematic upper limit for $M^{2}(b c d e)$ is given by

\begin{equation}
M^{2}_{\rm max}(b c d e) = \left\{
\begin{array}{lcccc}
\vspace{0.25cm}
\frac{(m_{A}^{2}-m_{B}^{2}) (m_{B}^{2}-m_{E}^{2})}{m_{B}^{2}} & \textrm{iff} & & \frac{m_{A}}{m_{E}} > & \frac{m_{B}^{2}}{m_{E}^{2}} \\
\vspace{0.25cm}
\frac{(m_{A}^{2} m_{C}^{2} - m_{B}^{2} m_{E}^{2}) (m_{B}^{2}-m_{C}^{2})}{m_{B}^{2} m_{C}^{2}} & \textrm{iff} & &  \frac{m_{A}}{m_{E}} < & \frac{m_{B}^{2}}{m_{C}^{2}} \\
\vspace{0.25cm}
\frac{(m_{A}^{2} m_{D}^{2} - m_{C}^{2} m_{E}^{2}) (m_{C}^{2}-m_{D}^{2})}{m_{C}^{2} m_{D}^{2}} & \textrm{iff} & &  \frac{m_{A}}{m_{E}} < & \frac{m_{C}^{2}}{m_{D}^{2}} , \\
\vspace{0.25cm}
\frac{(m_{A}^{2}-m_{D}^{2}) (m_{D}^{2}-m_{E}^{2})}{m_{D}^{2}} & \textrm{iff} & & \frac{m_{A}}{m_{E}} < & \frac{m_{D}^{2}}{m_{E}^{2}} \\
\vspace{0.25cm}
(m_{A}-m_{E})^{2}  & \textrm{otherwise} & \\
\end{array} \right.
\end{equation}

\noindent where  $m_{A}$, $m_{B}$, $m_{C}$, $m_{D}$, and $m_{E}$ are the masses of particles $A$, $B$, $C$, $D$, and $E$ \cite{Gjelsten:2005aw}.

\newpage
\bibliographystyle{h-physrev}
\bibliography{ms.bbl}
\newpage

\end{document}